\documentclass[prb,twocolumn,showpacs]{revtex4}
\usepackage{graphicx} 

\usepackage{amsmath,amsthm,amssymb,mathrsfs}
\usepackage{bm}

\begin{document}

\title{Joule heating in spin Hall geometry}

\author{Tomohiro Taniguchi}
 \affiliation{
 National Institute of Advanced Industrial Science and Technology (AIST), Spintronics Research Center, Tsukuba, Ibaraki 305-8568, Japan
 }

 \date{\today} 
 \begin{abstract}
  {
The theoretical formula for the entropy production rate in the presence of spin current is derived 
using the spin-dependent transport equation and thermodynamics. 
This theory is applicable regardless of the source of the spin current, 
for example, an electric field, a temperature gradient, or the Hall effect. 
It reproduces the result in a previous work on the dissipation formula  
when the relaxation time approximation is applied to the spin relaxation rate. 
By using the developed theory, 
it is found that the dissipation in spin Hall geometry has a contribution proportional to the square of the spin Hall angle. 
  }
 \end{abstract}

 \maketitle


Reducing the power consumption in spintronics devices is an important problem for 
both fundamental physics and practical applications. 
Therefore, the development of a theory of dissipation (heating) due to the spin current 
is an attractive topic in this field \cite{tulapurkar11,wegrowe00,sears11,taniguchi14}. 
In 2011, Tulapurkar and Suzuki \cite{tulapurkar11} investigated 
the dissipation in a one-dimensional ferromagnetic multilayer using the Boltzmann equation, 
where the spin current is driven by the electric field $E_{x}$ \cite{tulapurkar11}. 
In this case, the electric current density carried by spin-$\nu$ electrons ($\nu=\uparrow,\downarrow$ or $\pm$) 
is expressed as $J_{{\rm c},\nu}=\sigma_{\nu}E_{x}$, 
where $\sigma_{\nu}$ is the conductivity of the spin-$\nu$ electrons. 
They found an additional contribution proportional to $\beta^{2}/(1-\beta^{2})$ to 
the conventional Joule heating $\sigma E_{x}^{2}$, 
where $\sigma=\sigma_{\uparrow}+\sigma_{\downarrow}$ and $\beta=(\sigma_{\uparrow}-\sigma_{\downarrow})/\sigma$ 
are the total conductivity and its spin polarization, respectively. 


On the other hand, several methods of generating spin current, 
such as nonlocal spin-injection by diffusion \cite{johnson85,jedema01,kimura05}, 
spin pumping by ferromagnetic resonance \cite{silsbee79,mizukami02b,tserkovnyak02a}, 
the spin Hall effect due to spin-orbit interaction \cite{dyakonov71,hirsch99,kato04,ando08,kim12,taniguchi15}, 
the spin Seebeck effect due to heating \cite{uchida08,uchida10,slachter10,xiao10,adachi10}, 
and spin hydrodynamic generation by fluid dynamics \cite{matsuo13,takahashi16}, 
have recently been proposed theoretically and also demonstrated experimentally. 
The spin currents driven by these effects are not described by the Boltzmann equation used in Ref. \cite{tulapurkar11}. 
From this perspective, it is unclear whether the dissipation formula derived in a previous work \cite{tulapurkar11} is still applicable to these cases. 
Therefore, it is desirable to derive a dissipation formula of the spin current, 
that is independent of the explicit form of the source term (driving force) of the current. 


In this letter, we derive the theoretical formula of the entropy production rate 
for systems having spin-dependent transport properties 
by using the continuous equation of the spin current and thermodynamics. 
The derived formula is applicable regardless of the source of the spin current 
because it requires no explicit specification of the form of the current. 
We also show that the present formula reproduces the dissipation formula derived by Tulapurkar and Suzuki \cite{tulapurkar11} 
when the relaxation time approximation is applied to the spin relaxation rate. 
We apply the present formula to derive the dissipation in the spin Hall geometry, 
and find a contribution proportional to the square of the spin Hall angle. 


We first consider the spin-dependent transport of electrons in condensed matter, 
where the particle density $n_{\nu}$ and current density $\mathbf{j}_{\nu}$ of spin-$\nu$ electrons satisfy 
\begin{equation}
  \frac{\partial n_{\nu}}{\partial t}
  +
  \bm{\nabla}
  \cdot
  \mathbf{j}_{\nu}
  =
  -\varphi_{\nu},
  \label{eq:conservation_particle}
\end{equation}
where $\varphi_{\nu}$ is the spin relaxation rate. 
Because of the conservation law of the particle number, $\varphi_{\nu}$ should satisfy $\varphi_{\uparrow}+\varphi_{\downarrow}=0$. 
The particle current density $\mathbf{j}_{\nu}$ is related to the electric current density via 
$\mathbf{J}_{{\rm c},\nu}=-e\mathbf{j}_{\nu}$, 
where $e=|e|$ is the elementary charge. 
The total electric current density is $\mathbf{J}_{\rm c}=\mathbf{J}_{\rm c,\uparrow}+\mathbf{J}_{\rm c,\downarrow}$, 
whereas the spin current density is $\mathbf{J}_{\rm s}=[\hbar/(-2e)](\mathbf{J}_{\rm c,\uparrow}-\mathbf{J}_{\rm c,\downarrow})$. 
There are several sources of the current $\mathbf{J}_{\rm c}$ (or $\mathbf{J}_{{\rm c},\nu}$). 
For example, when an external electric field $\mathbf{E}$ drives the current, 
$\mathbf{J}_{\rm c}$ is given by $\sigma \mathbf{E}$. 
On the other hand, the current density is given by $\mathbf{J}_{\rm c}=\sigma S \bm{\nabla}T$ when a temperature gradient drives the current, 
where $S$ and $T$ are the Seebeck coefficient and the temperature, respectively. 
We emphasize that the following derivation of the entropy production rate formula 
is independent of the explicit form of $\mathbf{J}_{\rm c}$. 
In other words, the following calculation is applicable not only to 
the one-dimensional system with an electric field studied in Ref. \cite{tulapurkar11} 
but also to other systems. 


The total energy density $\mathscr{E}$ consists of the internal energy density $u$ and potential energy density $W$, 
where $W=-\sum_{\nu=\uparrow,\downarrow}en_{\nu}V$ with an electric voltage $V$. 
Therefore, the change in the total energy density $\mathscr{E}$ is 
\begin{equation}
  \frac{\partial \mathscr{E}}{\partial t}
  =
  \frac{\partial u}{\partial t}
  +
  \frac{\partial W}{\partial t}.
  \label{eq:conservation_energy}
\end{equation}
Similarly, the energy current density $\mathbf{j}_{E}$ is related to the internal energy current density $\mathbf{j}_{u}$ 
and the particle current density of the spin-$\nu$ electrons $\mathbf{j}_{\nu}$ via 
\begin{equation}
  \mathbf{j}_{E}
  =
  \mathbf{j}_{u}
  +
  \sum_{\nu=\uparrow,\downarrow}
  (-eV)
  \mathbf{j}_{\nu}.
  \label{eq:energy_current}
\end{equation}
The total energy density $\mathscr{E}$ and the energy current density $\mathbf{j}_{E}$ satisfy the energy conservation law 
\begin{equation}
  \frac{\partial \mathscr{E}}{\partial t}
  +
  \bm{\nabla}
  \cdot
  \mathbf{j}_{E}
  =
  0. 
\end{equation}
In the steady state, $\bm{\nabla}\cdot\mathbf{j}_{E}=0$. 
Then, we find that the Joule heating formula, $\bm{\nabla}\cdot\mathbf{j}_{u}=\mathbf{J}_{\rm c}\cdot\mathbf{E}$, 
is reproduced from Eq. (\ref{eq:energy_current}) and the conservation law of the particle number, $\sum_{\nu=\uparrow,\downarrow}\bm{\nabla}\cdot\mathbf{j}_{\nu}=0$, 
where the electric field is related to the electric voltage $V$ via $\mathbf{E}=-\bm{\nabla}V$. 


According to thermodynamics, the change in the internal energy density is \cite{kondepudi98} 
\begin{equation}
  \frac{\partial u}{\partial t}
  =
  T 
  \frac{\partial \mathfrak{S}}{\partial t}
  +
  \sum_{\nu=\uparrow,\downarrow}
  \mu_{\nu}
  \frac{\partial n_{\nu}}{\partial t},
  \label{eq:internal_energy}
\end{equation}
where $\mathfrak{S}$ is the entropy density, 
and $\mu_{\nu}$ is the chemical potential of the spin-$\nu$ electrons. 
The electrochemical potential is $\bar{\mu}_{\nu}=\mu_{\nu}-eV$. 
We also define the heat current density $\mathbf{j}_{q}$ as 
\begin{equation}
  \mathbf{j}_{q}
  =
  \mathbf{j}_{u}
  -
  \sum_{\nu=\uparrow,\downarrow}
  \mu_{\nu}
  \mathbf{j}_{\nu}.
\end{equation}
From Eq. (\ref{eq:energy_current}), the energy and heat current densities are related as 
\begin{equation}
  \mathbf{j}_{E}
  =
  \mathbf{j}_{q}
  +
  \sum_{\nu=\uparrow,\downarrow}
  \bar{\mu}_{\nu}
  \mathbf{j}_{\nu}.
  \label{eq:energy_heat_current}
\end{equation}
Using Eqs. (\ref{eq:conservation_particle})-(\ref{eq:energy_heat_current}), 
we find that 
\begin{equation}
  T 
  \frac{\partial \mathfrak{S}}{\partial t}
  =
  -\bm{\nabla}
  \cdot
  \mathbf{j}_{q}
  -
  \sum_{\nu=\uparrow,\downarrow}
  \mathbf{j}_{\nu}
  \cdot
  \bm{\nabla}
  \bar{\mu}_{\nu}
  +
  \sum_{\nu=\uparrow,\downarrow}
  \bar{\mu}_{\nu}
  \varphi_{\nu}.
\end{equation}
The equation can be rewritten as 
\begin{equation}
  \frac{\partial \mathfrak{S}}{\partial t}
  +
  \bm{\nabla}
  \cdot
  \frac{\mathbf{j}_{q}}{T}
  =
  \mathbf{j}_{q}
  \cdot
  \bm{\nabla}
  \frac{1}{T}
  -
  \frac{1}{T}
  \sum_{\nu=\uparrow,\downarrow}
  \mathbf{j}_{\nu}
  \cdot
  \bm{\nabla}
  \bar{\mu}_{\nu}
  +
  \frac{1}{T}
  \sum_{\nu=\uparrow,\downarrow}
  \bar{\mu}_{\nu}
  \varphi_{\nu}.
  \label{eq:entropy_production}
\end{equation}
The right-hand side of Eq. (\ref{eq:entropy_production}) is the entropy production rate \cite{kondepudi98} 
including the spin degree of freedom. 
The last term of Eq. (\ref{eq:entropy_production}) appears from the spin-dependent relaxation rate 
and has not appeared in the classical theory of dissipation, 
which does not take the spin current into account \cite{abrikosov88}. 
Note that the derivation of Eq. (\ref{eq:entropy_production}) is independent of the explicit form of the particle current density. 
Therefore, Eq. (\ref{eq:entropy_production}) is general and valid for any type of spin current source (driving force), as emphasized earlier. 


Tulalpurkar and Suzuki \cite{tulapurkar11} derived 
a different form of the entropy production rate formula 
in terms of the electrochemical potential $\bar{\mu}=(\bar{\mu}_{\uparrow}+\bar{\mu}_{\downarrow})/2$ 
and the spin accumulation $\delta\mu=(\bar{\mu}_{\uparrow}-\bar{\mu}_{\downarrow})/2$ 
by using the Boltzmann equation. 
Here, we show that the result in Ref. \cite{tulapurkar11} is reproduced from Eq. (\ref{eq:entropy_production}) 
by applying the relaxation time approximation \cite{valet93} to the spin relaxation rate given by 
\begin{equation}
  \varphi_{\nu}
  =
  \frac{n_{\nu}}{2 \tau_{\rm sf}^{\nu}}
  -
  \frac{n_{-\nu}}{2 \tau_{\rm sf}^{-\nu}},
  \label{eq:spin_relaxation_rate}
\end{equation}
where $\tau_{\rm sf}^{\nu}$ is the spin-flip relaxation time from the spin-$\nu$ state to the opposite spin state. 
The nonequilibrium particle density, $n_{\nu}$, is related to the density of state $\mathcal{N}_{\nu}$ at the Fermi level $\varepsilon_{\rm F}$ 
via $n_{\nu}=\mathcal{N}_{\nu}(\mu_{\nu}-\varepsilon_{\rm F})$. 
The detailed balance, $\mathcal{N}_{\nu}/\tau_{\rm sf}^{\nu}=\mathcal{N}_{-\nu}/\tau_{\rm sf}^{-\nu}$, is satisfied 
in the steady state \cite{hershfield97}. 
Using Eq. (\ref{eq:spin_relaxation_rate}) and these relations, we find that 
\begin{equation}
  \sum_{\nu=\uparrow,\downarrow}
  \bar{\mu}_{\nu}
  \varphi_{\nu}
  =
  \frac{(1-\beta^{2})}{e^{2}\rho \ell^{2}}
  \left(
    \delta 
    \mu
  \right)^{2}.
  \label{eq:Valet_Fert}
\end{equation}
The spin diffusion length $\ell$ is defined as 
\begin{equation}
\begin{split}
  \frac{1}{\ell^{2}}
  &=
  \frac{e^{2}}{2}
  \left(
    \frac{\mathcal{N}_{\uparrow}}{\sigma_{\uparrow}\tau_{\rm sf}^{\uparrow}}
    +
    \frac{\mathcal{N}_{\downarrow}}{\sigma_{\downarrow}\tau_{\rm sf}^{\downarrow}}
  \right)
\\
  &=
  \frac{1}{2}
  \left(
    \frac{1}{D_{\uparrow}\tau_{\rm sf}^{\uparrow}}
    +
    \frac{1}{D_{\downarrow}\tau_{\rm sf}^{\downarrow}}
  \right),
\end{split}
\end{equation}
where the diffusion coefficient $D_{\nu}$ is related to the conductivity of the spin-$\nu$ electrons, $\sigma_{\nu}$, 
by the Einstein relation, $\sigma_{\nu}=e^{2}D_{\nu}\mathcal{N}_{\nu}$. 
We also find from Eqs. (\ref{eq:conservation_particle}) and (\ref{eq:spin_relaxation_rate}) that 
\begin{equation}
\begin{split}
  \bm{\nabla}
  \cdot
  \left(
    \mathbf{j}_{\uparrow}
    -
    \mathbf{j}_{\downarrow}
  \right)
  &=
  -\frac{(1-\beta^{2}) \sigma}{e^{2}\ell^{2}}
  \delta 
  \mu.
  \label{eq:VF_1}
\end{split}
\end{equation}
Substituting Eq. (\ref{eq:VF_1}) into (\ref{eq:Valet_Fert}), 
we obtain 
\begin{equation}
  \sum_{\nu=\uparrow,\downarrow}
  \bar{\mu}_{\nu}
  \varphi_{\nu}
  =
  -\left[
    \bm{\nabla}
    \cdot
    \left(
      \mathbf{j}_{\uparrow}
      -
      \mathbf{j}_{\downarrow}
    \right)
  \right]
  \delta
  \mu.
  \label{eq:VF_2}
\end{equation}
Substituting Eq. (\ref{eq:VF_2}) into Eq. (\ref{eq:entropy_production}) and using the conservation law of the particle current, 
$\bm{\nabla}\cdot(\mathbf{j}_{\uparrow}+\mathbf{j}_{\downarrow})=0$, 
we find that 
\begin{equation}
  \frac{\partial \mathfrak{S}}{\partial t}
  +
  \bm{\nabla}
  \cdot
  \frac{\mathbf{j}_{q}}{T}
  =
  \mathbf{j}_{q}
  \cdot
  \bm{\nabla}
  \frac{1}{T}
  -
  \frac{1}{T}
  \bm{\nabla}
  \cdot
  \left[
    \left(
      \mathbf{j}_{\uparrow}
      +
      \mathbf{j}_{\downarrow}
    \right)
    \bar{\mu}
  \right]
  -
  \frac{1}{T}
  \bm{\nabla}
  \cdot
  \left[
    \left(
      \mathbf{j}_{\uparrow}
      -
      \mathbf{j}_{\downarrow}
    \right)
    \delta \mu
  \right].
  \label{eq:entropy_production_VF}
\end{equation}
The second and third terms on the right-hand side of Eq. (\ref{eq:entropy_production_VF}) include 
the electric and spin current densities, $\mathbf{J}_{\rm c}=-e(\mathbf{j}_{\uparrow}+\mathbf{j}_{\downarrow})$ 
and $\mathbf{J}_{\rm s}=(\hbar/2)(\mathbf{j}_{\uparrow}-\mathbf{j}_{\downarrow})$, respectively. 
Equation (\ref{eq:entropy_production_VF}) is identical to the dissipation formula in Ref. \cite{tulapurkar11} 
derived using the Boltzmann equation. 
We denote the right hand side of Eq. (\ref{eq:entropy_production_VF}) as $\Sigma_{V}$ for convenience, that is,
\begin{equation}
  \Sigma_{V}
  =
  \mathbf{j}_{q}
  \cdot
  \bm{\nabla}
  \frac{1}{T}
  -
  \frac{1}{T}
  \bm{\nabla}
  \cdot
  \left[
    \left(
      \mathbf{j}_{\uparrow}
      +
      \mathbf{j}_{\downarrow}
    \right)
    \bar{\mu}
  \right]
  -
  \frac{1}{T}
  \bm{\nabla}
  \cdot
  \left[
    \left(
      \mathbf{j}_{\uparrow}
      -
      \mathbf{j}_{\downarrow}
    \right)
    \delta \mu
  \right].
  \label{eq:entropy_production_bulk}
\end{equation}
We emphasize that $\Sigma_{V}$ is the bulk entropy production rate. 
Similarly, the interface entropy production rate is given by \cite{tulapurkar11} 
\begin{equation}
  \Sigma_{A}
  =
  \hat{\mathbf{n}}
  \cdot
  \mathbf{j}_{q}
  \Delta
  \frac{1}{T}
  -
  \frac{1}{T}
  \hat{\mathbf{n}}
  \cdot
  \left[
    \left(
      \mathbf{j}_{\uparrow}
      +
      \mathbf{j}_{\downarrow}
    \right)
    \Delta
    \bar{\mu}
  \right]
  -
  \frac{1}{T}
  \hat{\mathbf{n}}
  \cdot
  \left[
    \left(
      \mathbf{j}_{\uparrow}
      -
      \mathbf{j}_{\downarrow}
    \right)
    \Delta
    \delta
    \mu
  \right],
  \label{eq:entropy_production_interface}
\end{equation}
where $\hat{\mathbf{n}}$ is the unit vector normal to the interface, 
and $\Delta(1/T)$, $\Delta \bar{\mu}$, and $\Delta \delta \mu$ are 
the differences in $1/T$, $\bar{\mu}$, and $\delta\mu$ at the interface, respectively. 
We assume that the heat, electric, and spin currents are continuous at the interface. 
The total entropy production rate is given by the sum of Eqs. (\ref{eq:entropy_production_bulk}) and (\ref{eq:entropy_production_interface}) 
integrated over the volume and interfaces, respectively. 
For example, let us consider the total entropy production rate in the current-perpendicular-to-plane spin valve 
consisting of two ferromagnets, F$_{1}$ and F${}_{2}$, and a nonmagnet N. 
Assuming that the electric current flows along the $x$ direction, 
and summing the bulk entropy production rates $\Sigma_{V}$ in the F${}_{1}$, N, and F${}_{2}$ layers 
and the interface entropy production rates $\Sigma_{A}$ at the F${}_{1}$/N and N/F${}_{2}$ interfaces, 
the total entropy production per unit area per unit time becomes 
\begin{equation}
\begin{split}
  &
  \int_{-\infty}^{\rm F_{1}/N} 
  dx 
  \Sigma_{V} 
  + 
  \Sigma_{A}({\rm F}_{1}/{\rm N})
  +
  \int_{\rm N} 
  dx 
  \Sigma_{V}
  +
  \Sigma_{A}({\rm N}/{\rm F}_{2})
  +
  \int_{\rm N/F_{2}}^{\infty}
  dx 
  \Sigma_{V}
\\
  &=
  \frac{J_{\rm c}}{eT}
  \left[
    \bar{\mu}(\infty)
    -
    \bar{\mu}(-\infty)
  \right],
  \label{eq:entropy_production_TS}
\end{split}
\end{equation}
as derived in Ref. \cite{tulapurkar11}, 
where the system is connected to the electrodes at $x \to \pm \infty$. 
We assume that $\lim_{x \to \pm \infty}\delta\mu=0$ and 
the temperature profile is uniform. 
Note that $\bar{\mu}(\infty)-\bar{\mu}(-\infty)$ in Eq. (\ref{eq:entropy_production_TS}) is the potential difference 
between the electrodes, which drives the electric current. 
The entropy production rate originating from the spin Hall effect is calculated in a similar way, as shown below. 
Similarly, when the system is connected to a heat bath, 
the entropy production rate will be related to the temperature difference between the boundaries, 
as implied by the first terms of Eqs. (\ref{eq:entropy_production_bulk}) and (\ref{eq:entropy_production_interface}). 
Another example of the previously reported entropy production rate is that in spin pumping \cite{taniguchi14}. 
In this case, a ferromagnetic(F)/nonmagnetic(N) interface with the ferromagnet in resonance plays 
a role similar to the electrodes in the spin valve because the spin current is driven from this interface. 
Then, as the result of an integral similar to that in Eq. (\ref{eq:entropy_production_TS}), 
the total entropy production rate per unit area becomes $[J_{\rm s}/(\hbar T/2)](\delta\mu_{\rm N}-\delta\mu_{\rm F})$ as studied in Ref. \cite{taniguchi14}, 
where $J_{\rm s}$, $\delta\mu_{\rm N}$, and $\delta\mu_{\rm F}$ are 
the total spin current pumped at the F/N interface and the spin accumulations at the interfaces of the N and F layers, respectively. 
We note that $J_{\rm s}$ and $\delta\mu_{\rm N}-\delta\mu_{\rm F}$ in this expression correspond to 
$J_{\rm c}$ and $\bar{\mu}(\infty)-\bar{\mu}(-\infty)$, respectively, in Eq. (\ref{eq:entropy_production_TS}). 
In both cases, the total entropy production rate per is given by 
the current multiplied by the energy (potential difference) driving this current, as expected from thermodynamics \cite{kondepudi98}. 

\begin{figure}
\centerline{\includegraphics[width=1.0\columnwidth]{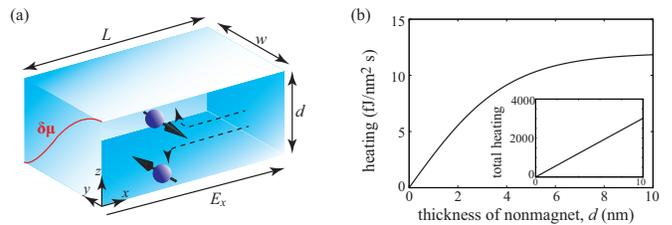}}
\caption{
        (a) Schematic view of the spin Hall system. 
        (b) Dependence of the second term of Eq. (\ref{eq:heating_SHE}), divided by the cross-sectional area $Lw$, on the thickness of the nonmagnet $d$. 
            Inset shows the total dissipation [sum of the first and second terms of Eq. (\ref{eq:heating_SHE})]. 
         \vspace{-3ex}}
\label{fig:fig1}
\end{figure}

Now let us apply the above formula to estimate the dissipation due to the spin current 
by considering the spin Hall system as an example. 
The system under consideration is shown in Fig. \ref{fig:fig1}, 
where an electric field, $E_{x}$, applied to a nonmagnet along the $x$ direction 
drives the spin-dependent electric current $\mathbf{J}_{{\rm c},\nu}$. 
The length along the $x$ direction, the width along the $y$ direction, and the thickness along the $z$ direction are 
denoted as $L$, $w$, and $d$, respectively. 
We assume a homogeneous temperature profile, that is, $\bm{\nabla}T=\bm{0}$, for simplicity. 
The spin-orbit interaction in the nonmagnet scatters the electrons in the transverse direction, 
where the scattering direction is spin-dependent. 
In the spin Hall geometry, the electric current densities carried by the spin-up and spin-down electrons are given by 
\begin{equation}
  \mathbf{J}_{{\rm c},\uparrow}
  =
  \frac{\sigma_{\rm N}}{2e}
  \bm{\nabla}
  \bar{\mu}_{\uparrow}
  +
  \frac{\vartheta \sigma_{\rm N}}{2e}
  \hat{\mathbf{s}}
  \times
  \bm{\nabla}
  \bar{\mu}_{\uparrow},
  \label{eq:spin_up}
\end{equation}
\begin{equation}
  \mathbf{J}_{{\rm c},\downarrow}
  =
  \frac{\sigma_{\rm N}}{2e}
  \bm{\nabla}
  \bar{\mu}_{\downarrow}
  -
  \frac{\vartheta \sigma_{\rm N}}{2e}
  \hat{\mathbf{s}}
  \times
  \bm{\nabla}
  \bar{\mu}_{\downarrow},
  \label{eq:spin_down}
\end{equation}
where $\sigma_{\rm N}$, $\vartheta$, and $\hat{\mathbf{s}}$ are the conductivity of the nonmagnet, 
the spin Hall angle, and the unit vector pointing in the direction of the spin polarization, respectively. 
The electric current density and 
spin current density are 
\begin{equation}
  \mathbf{J}_{\rm c}
  =
  \frac{\sigma_{\rm N}}{e}
  \bm{\nabla}
  \bar{\mu}
  +
  \frac{\vartheta \sigma_{\rm N}}{e}
  \hat{\mathbf{s}}
  \times
  \bm{\nabla}
  \delta\mu,
  \label{eq:electric_current_SHE}
\end{equation}
\begin{equation}
  \mathbf{J}_{\rm s}
  =
  -\frac{\hbar \sigma_{\rm N}}{2e^{2}}
  \bm{\nabla}
  \delta
  \mu
  -
  \frac{\hbar \vartheta \sigma_{\rm N}}{2e^{2}}
  \hat{\mathbf{s}}
  \times
  \bm{\nabla}
  \bar{\mu}. 
  \label{eq:spin_current_SHE}
\end{equation}
It was shown \cite{takahashi08} that the relaxation time approximation is applicable to the diffusion equation of the spin accumulation $\delta\mu$. 
Thus, Eq. (\ref{eq:entropy_production_VF}) can be used to derive the entropy production rate formula in the spin Hall geometry. 




We assume that the width $w$ is sufficiently large, 
and the spatial variations of $\bar{\mu}$ and $\delta\mu$ along the $y$ direction are negligible, as in the case of the experiments. 
Then, the spatial derivatives of $\bar{\mu}$ and $\delta\mu$ are given by 
$\bm{\nabla}(\bar{\mu}/e)=E_{x}\mathbf{e}_{x}+\partial_{z}(\bar{\mu}/e)\mathbf{e}_{z}$ and 
$\bm{\nabla}(\delta\mu/e)=\partial_{z}(\delta\mu/e)\mathbf{e}_{z}$, respectively. 
Applying the open circuit conditions of the electric current along the $z$ direction, 
we find that $\partial_{z}\bar{\mu}=0$. 
Thus, the electrochemical potential is $\bar{\mu}=eE_{x}x$. 
On the other hand, by solving the diffusion equation of the spin accumulation $\delta\mu$ 
with the open circuit conditions of the spin current and Eq. (\ref{eq:spin_current_SHE}), 
the solution of $\delta\mu$ is given by 
\begin{equation}
  \delta
  \mu
  =
  \frac{-e \vartheta E_{x} \ell}{\sinh(d/\ell)}
  \left[
    \cosh
    \left(
      \frac{z-d}{\ell}
    \right)
    -
    \cosh
    \left(
      \frac{z}{\ell}
    \right)
  \right],
  \label{eq:spin_accumulation_solution}
\end{equation}
where $\ell$ is the spin diffusion length of the nonmagnet. 
Note that the spin accumulation is related to the electric current along the $x$ direction as 
$\mathbf{e}_{x}\cdot\mathbf{J}_{\rm c}=\sigma_{\rm N}E_{x}+(\vartheta \sigma_{\rm N}/e)\partial_{z}\delta\mu$. 
Let us denote Eq. (\ref{eq:entropy_production_bulk}) integrated over the volume of the nonmagnet 
and multiplied by the temperature 
as $\partial \mathcal{Q}_{V}/\partial t = \int dx dy dz T \Sigma_{V}$, 
which is the dissipation (heating) per unit time in this system. 
Using the above result, $\bm{\nabla}(\bar{\mu}/e)=E_{x}\mathbf{e}_{x}$, 
and applying the open-circuit condition to the spin current, 
we find that 
\begin{equation}
\begin{split}
  \frac{\partial \mathcal{Q}_{V}}{\partial t}
  &=
  \int dx dy dz
  \left(
    \sigma_{\rm N}
    E_{x}
    +
    \frac{\vartheta \sigma_{\rm N}}{e}
    \partial_{z}
    \delta
    \mu
  \right)
  E_{x}.
  \label{eq:volume_integral}
\end{split}
\end{equation}
Substituting Eq. (\ref{eq:spin_accumulation_solution}) into Eq. (\ref{eq:volume_integral}), 
we find that 
\begin{equation}
  \frac{\partial \mathcal{Q}_{V}}{\partial t}
  =
  \sigma_{\rm N}
  E_{x}^{2}
  L w d 
  +
  2 \vartheta^{2}
  \sigma_{\rm N}
  E_{x}^{2}
  Lw \ell 
  \tanh
  \left(
    \frac{d}{2 \ell}
  \right). 
  \label{eq:heating_SHE}
\end{equation}
The first term of Eq. (\ref{eq:heating_SHE}) is the conventional Joule heating, 
which is the product of the electric current density $\sigma_{\rm N}E_{x}$ and the electric field $E_{x}$ 
integrated over the volume. 
On the other hand, the second term represents the contribution 
of the inverse spin Hall effect to the dissipation. 
This term is proportional to the square of the spin Hall angle 
and becomes saturated for a thickness sufficiently thicker than the spin diffusion length. 
Figure \ref{fig:fig1}(b) shows an example of the dependence of the second term of Eq. (\ref{eq:heating_SHE}) 
on the thickness of the nonmagnet, $d$. 
In the figure, the second term of Eq. (\ref{eq:heating_SHE}), shown on the vertical axis, is 
divided by the cross-sectional area in the $xy$ plane ($Lw$), 
that is, $2 \vartheta^{2} \sigma_{\rm N}E_{x}^{2} \ell \tanh[d/(2 \ell)]$. 
The values of the parameters are $\rho_{\rm N}=1/\sigma_{\rm N}=3000$ $\Omega$nm, $\ell=2$ nm, 
and $|\vartheta|=0.1$, which are typical values found in experiments 
for nonmagnetic heavy metals \cite{ando08,kim12,niimi14,torrejon15}. 
The electric field is $E_{x}=J_{\rm c0}/\sigma_{\rm N}$, where the current density is $J_{\rm c0}=10^{6}$ A/cm${}^{2}$. 
As shown, this dissipation is on the order of 10 fJ/(nm${}^{2}$s), which is comparable to that found in the other system \cite{taniguchi14}. 
Because the spin Hall angle is small, this dissipation is much smaller than the total dissipation, 
which is shown in the inset of Fig. \ref{fig:fig1}(b). 


In conclusion, 
we developed a comprehensive theory of dissipation due to a spin current 
by using the continuous equation of the spin current and thermodynamics. 
A formula for the entropy production rate that is applicable regardless of the spin current was derived. 
The previous work by Tulapurkar and Suzuki \cite{tulapurkar11} is reproduced when the relaxation time approximation for the spin relaxation 
is applied to the present theory. 
We also found an additional term proportional to the square of the spin Hall angle contributing to 
the conventional Joule heating in the spin Hall geometry. 


The author is grateful to Mark D. Stiles and Wayne M. Saslow for valuable discussion. 
The author is also thankful to Shinji Yuasa, Hitoshi Kubota, Kay Yakushiji, Akio Fukushima, Yoshishige Suzuki, Takayuki Nozaki, Makoto Konoto, Takehiko Yorozu, 
Satoshi Iba, Yoichi Shiota, Sumito Tsunegi, Atsushi Sugihara, Takahide Kubota, Shinji Miwa, Hiroki Maehara, and Ai Emura 
for their support and encouragement. 
This work was supported by JSPS KAKENHI Grant-in-Aid for Young Scientists (B) 16K17486. 




\end{document}